\documentclass[prl,aps,twocolumn,groupedaddress,floats,showpacs,final]{revtex4-1}
\usepackage{graphicx}
\usepackage{dcolumn}
\usepackage{bm}
\usepackage{color}
\usepackage[caption=false]{subfig}

\definecolor{DarkGreen}{rgb}{0,0.7,0.08} 
\definecolor{Grey}{rgb}{0.5,0.5,0.5}
\definecolor{Red}{rgb}{0.8,0.3,0.3}
\definecolor{Green}{rgb}{0.1,0.8,0.1}
\definecolor{Blue}{rgb}{0.1,0.1,0.8}
\newcommand{\oldtext}[1]{}

\begin{document}

\title{Quasi-2D liquid  $^3$He}
\author{Michele Ruggeri$^{1}$,  Saverio Moroni$^1$, and Massimo Boninsegni$^{2}$}
\affiliation{$^1$ {SISSA Scuola Internazionale Superiore di Studi Avanzati and DEMOCRITOS National 
Simulation Center,
Istituto Officina dei Materiali del CNR Via Bonomea 265, I-34136, Trieste, Italy}}
\affiliation{$^2$ Department of Physics, University of Alberta, Edmonton, Alberta, Canada, T6G 2E1}
\date{\today}                                           

\begin{abstract}
Quantum Monte Carlo simulations at zero temperature of an ensemble of  $^3$He  atoms adsorbed on Mg and Alkali substrates  yield strong evidence of a thermodynamically stable liquid $^3$He monolayer on all Alkali substrates, with the possible exception of Li. The effective 2D density is $\theta\approx 0.02$ \AA$^{-2}$ on Na, making it the lowest density liquid   in Nature.  Its existence is underlain by zero-point atomic motion perpendicular to the substrate, whose effect is softening the short-range repulsion of the  helium interatomic potential. The monolayer films should turn superfluid at a temperature $T_c\sim 1$ mK. No liquid film is predicted to form on Mg, or on stronger substrates such as graphite.

\end{abstract}

\maketitle

Decades of experimental and theoretical investigation have yielded evidence that the behaviour of sufficiently thin films of $^4$He closely mimics that  theoretically predicted for a system strictly confined to two spatial dimensions (2D), atomic zero-point motion in the direction perpendicular to the substrate notwithstanding.  For example, the superfluid transition in thin liquid $^4$He films has been observed \cite{bishop,agnolet,tulimieri,luhman}  to conform to the  2D Kosterlitz-Thouless paradigm  \cite{kosterlitz}.  
The effect of substrate corrugation and/or roughness is generally limited to the appearance of insulating phases, either disordered or  crystalline, registered with the underlying substrate, at commensurate coverages.  Thus, thin films of $^4$He constitute an ideal quasi-2D  system, on which fundamental properties of strongly interacting Bose fluids in two dimensions can be investigated experimentally.
\\ \indent
The lighter isotope of helium ($^3$He) could in principle play the same role for Fermi systems, as $^3$He atoms are composite spin-1/2 particles. However, while $^3$He forms registered solid monolayers on corrugated substrates, with atoms sitting at preferential adsorption sites, the conventional wisdom is that on a substrate  $^3$He will only form  a ``thick" liquid film, whose physical properties are essentially those of the bulk liquid phase. The theoretical explanation is that, unlike $^4$He, whose 2D equilibrium phase at temperature $T$=0 is a liquid with a binding energy of approximately 1 K per atom \cite{whitlock}, no self-bound liquid is deemed to exist  in 2D for $^3$He \cite{campbell,miller,grau,nava}.
\\ \indent
A significant amount of experimental work has focused on a possible quasi-2D liquid phase of highly dilute $^3$He, floating atop a thin $^4$He film adsorbed on different substrates \cite{gasparini,valles,csathy}. While some of these studies have yielded some evidence of it,  consensus 
is still lacking as to whether the results unambiguously point to the existence of a thermodynamically stable quasi-2D $^3$He liquid \cite{smart}. Furthermore, interesting as the physics of a film of a mixture of the two isotopes of helium undoubtedly is \cite{hallock2}, the physical environment experienced by a $^3$He atom moving on top of (or through) a superfluid $^4$He film cannot be regarded as fully equivalent to that of a planar surface, as the $^3$He atom can scatter off surface excitations (ripplons) of the underlying $^4$He film. Thus, the question remains of whether a substrate exists, 
on which $^3$He by itself could condense into a quasi-2D liquid. 
\\ \indent
Recent experiments \cite{sato} suggest that this may be the case for $^3$He adsorbed directly on graphite, specifically the apparent formation of liquid puddles in an adsorbed monolayer, at a coverage $\lesssim$ 0.01 \AA$^{-2}$, i.e., far less than that corresponding to a registered solid monolayer. In this case too, however, an unambiguous interpretation of the experimental evidence is complicated by the expected significant role played by surface imperfection and/or inhomogeneities \cite{smart}.
\\ \indent
In this Letter, we provide theoretical evidence that $^3$He will form quasi-2D liquid monolayers at $T$=0 on all alkali substrates with the possible exception of Li,  with a coverage of approximately 0.03 \AA$^{-2}$ on K, Rb and Cs, and 0.02 \AA$^{-2}$ on Na. 
We base this statement on atomic energetics computed by  Quantum Monte Carlo (QMC) simulations 
of a realistic model of a $^3$He adsorbate. We have considered six {\it weakly attractive} substrates, namely Mg, Li, Na, K, Rb and Cs (listed in order of decreasing attractiveness). 
Our findings are that a liquid monolayer will only form on substrates that are sufficiently weak, i.e., all the alkali ones with the possible exception of Li, which appears to be a borderline case. On the stronger Mg substrate, no thin $^3$He film will form. In all predicted thermodynamically stable liquid phases, zero-point motion in the direction perpendicular to the substrate is significant, and has the effect of softening the hard-core repulsion of the interatomic $^3$He pair potential at short distances, in turn allowing for a liquid phase to exist.
\\ \indent
Our system of interest is modeled as an ensemble of $N$ $^3$He atoms, ($N/2$ of either spin component, i.e., the system is unpolarized) regarded as point particles,  moving in the presence of an infinite, smooth planar substrate (positioned at $z=0$). The system is enclosed in a vessel shaped as a parallelepiped, with periodic boundary conditions in all directions. The length of the simulation cell in the direction perpendicular to the substrate $z$ is taken to be large enough (100 \AA) to make the boundary conditions in that direction immaterial. The nominal $^3$He coverage is $\theta=N/A$, $A$ being the  area of the substrate. The quantum-mechanical many-body  Hamiltonian  is the following:
\begin{equation}\label{one}
\hat H = -{\hbar^2\over 2m}\sum_{i=1}^N \nabla_i^2 + \sum_{i<j} V(r_{ij}) +\sum_{i=1}^N U(z_i)
\end{equation}
where $m$ is the $^3$He atomic mass, $V$ is the potential describing the interaction between two helium atoms,  only depending on their relative distance, and $U$ is the potential describing the interaction of a $^3$He atom with the substrate,  also depending only on the distance of the atom from the substrate. We use the accepted Aziz potential \cite{aziz79} to describe the interaction of two $^3$He atoms. 
For each of the substrates studied here, the $U$ term  in Eq. (\ref{one}) is a potential proposed by Chizmeshya, Cole and Zaremba \cite{chi98}. The assumption of a smooth, planar substrate is clearly an important one; its justification is provided by the relative weakness of the substrates considered here.
We do not consider here stronger substrates \cite{sato}, where such an assumption might be inaccurate.
\\ \indent
We investigate the ground state of (\ref{one}) by means of a QMC technique, specifically Diffusion Monte Carlo (DMC), which 
projects the lowest-energy component out of an initial trial wave function $\Psi_T$. This is a well-established methodology \cite{umrigar}, which has been utilized to study  ground state properties of a wide variety of quantum many-body systems. The trial wave function that we utilized in this work has the form $\Psi=F\Phi$, where, using standard notation (see, for instance, Refs. \cite{sklc,sklc2})
\begin{equation}
F = {\rm exp}\biggl[{-\frac{1}{2}\sum_{i<j=1}^N u_{ij}-\frac{1}{2}\sum_{i<j<k=1}^N w_{ijk}})\biggr ]
\end{equation}
and $\Phi\equiv \{\prod_{i=1}^Nf(z_i)\}\ \phi(\uparrow)\phi(\downarrow)$, $\phi(\alpha)$ ($\alpha=\uparrow,\downarrow$) given by
\begin{equation}\label{slater}
\phi(\alpha) = {\rm Det}_{ij}\ \biggl \{ {\rm exp}\biggl [i{\bf k}_i\cdot\biggl ({\bf s}_j+\sum_{k\ne j}^N \eta(s_{kj}){\bf s}_{kj}\biggl )\biggr ]
\biggr \}
\end{equation}
where ${\bf s}_i\equiv (x_i,y_i)$, the wave vectors ${\bf k}_i$ span the $N/2$ allowed momenta of a 2D Fermi sea; in Eq. (\ref{slater}), $i$ and $j$ range from 1 to $N/2$ ($N/2+1$ to $N$) for $\alpha=\uparrow (\downarrow)$. The two-, three-body and backflow correlation functions $u$, $w$ and $\eta$, as well as the single-particle orbital $f(z)$ are optimized based on the procedure described in Ref. \cite{fantoni}. \\
We make use of the well-known ``fixed-node" approximation (FNA) in order to circumvent the inevitable sign problem that affects any fermion QMC method. We are mainly interested in the energy per $^3$He atom  computed as a function of coverage, and the FNA has been shown to provide accurate, variational energy estimates for bulk 3D $^3$He \cite{holzmann}. 
\begin{figure}[h]
\includegraphics[scale=0.36]{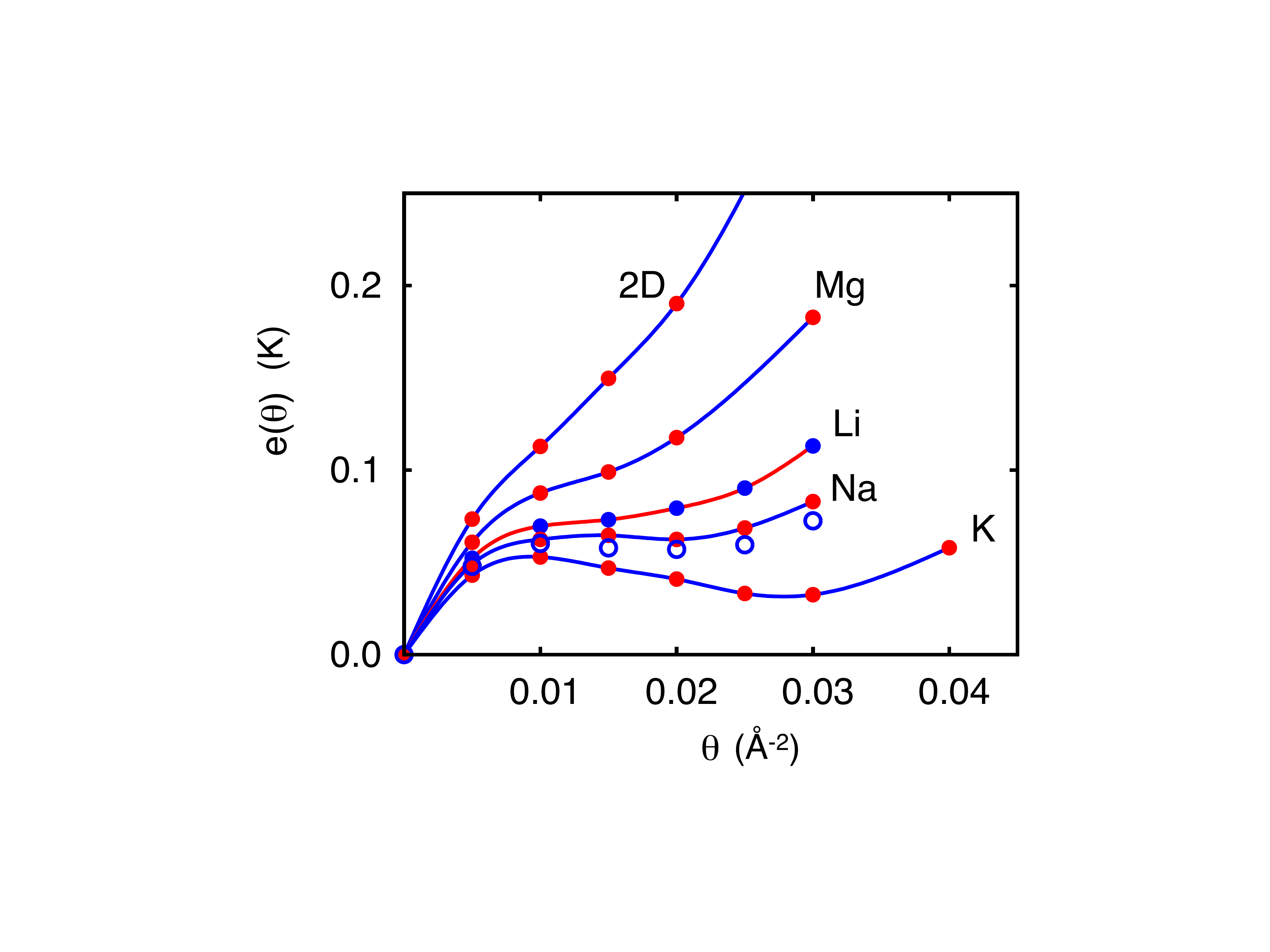}
\caption{(Color online)  Ground state energy per atom $e(\theta)$ computed by DMC for a $^3$He film adsorbed on different substrates, as a function of coverage $\theta$. The binding energy of a single atom on the given substrate is taken as the reference (zero) value for $e(\theta)$. Open symbols represent a second set of data for the Li substrate, obtained by applying estimates for the finite-size, time step, population control and fixed-node approximation errors. Solid lines are spline interpolations. The line for Li is colored differently, to indicate that the system is borderline for formation of a liquid layer. The data labelled ``2D" refer to strictly 2D $^3$He. Results for Cs and Rb substrates are indistinguishable from those on K, within the statistical errors of the calculations. Statistical errors are smaller than the size of the symbols.}
\label{f1}
\end{figure}
The simulations carried out in this work, for which results are shown, are for a system of $N$=26 $^3$He atoms; the DMC time step is $10^{-3}$ K$^{-1}$, and the population comprises 2,000 random walkers. We discuss systematic errors affecting our calculation after illustrating our results.
\\ \indent
Figure \ref{f1} shows the ground state energy per atom $e(\theta)$, computed by DMC as a function of coverage on the different substrates considered here. In all cases, the reference energy is taken to be the binding energy of a single $^3$He atom on the given substrate \cite{note2}. Also shown for comparison is the result for the strictly 2D system. 
The existence of a thermodynamically stable quasi-2D liquid phase is signalled by a local minimum of $e(\theta)$. Specifically, although the absolute minimum of the energy per particle remains at zero coverage  just like for purely 2D $^3$He, on increasing the chemical potential one observes a first order phase transition from a low density gas to a {\it liquid} of density close to (slightly greater than) that at which the  minimum is located.\\ One can clearly see that, starting from the strictly 2D case, and as the substrate is rendered weaker, i.e., atomic motion in the direction perpendicular to the substrate becomes more significant, the curve $e(\theta)$ bends downward. On Mg, namely the most attractive of the six substrates, $e(\theta)$ remains monotonically increasing,   i.e., no stable liquid film forms. The corresponding curve for any substrate more attractive than Mg (e.g., graphite), lies between that for Mg and the one labelled 2D in Figure \ref{f1}, implying  no liquid $^3$He film on graphite as well, in conflict with the claim of Ref. \cite{sato}, under the assumption of a smooth substrate. \\ 
\indent
The  downward bending of $e(\theta)$ becomes increasingly  pronounced on Li, Na and K, and a clear minimum can be observed on the last two substrates. On K, it is located at $\theta$=0.03 \AA$^{-2}$, whereas on Na at a coverage $\theta$=0.02 \AA$^{-2}$.  On Li, which is the weakest substrate on which $^4$He will form a superfluid monolayer \cite{toigo}, our numerical data show no evidence of a stable liquid $^3$He film, although, as discussed below, Li may be a borderline case, and we cannot exclude a fragile liquid phase at a coverage close to 0.015 \AA$^{-2}$.  
\begin{figure}[h]
\includegraphics[scale=0.36]{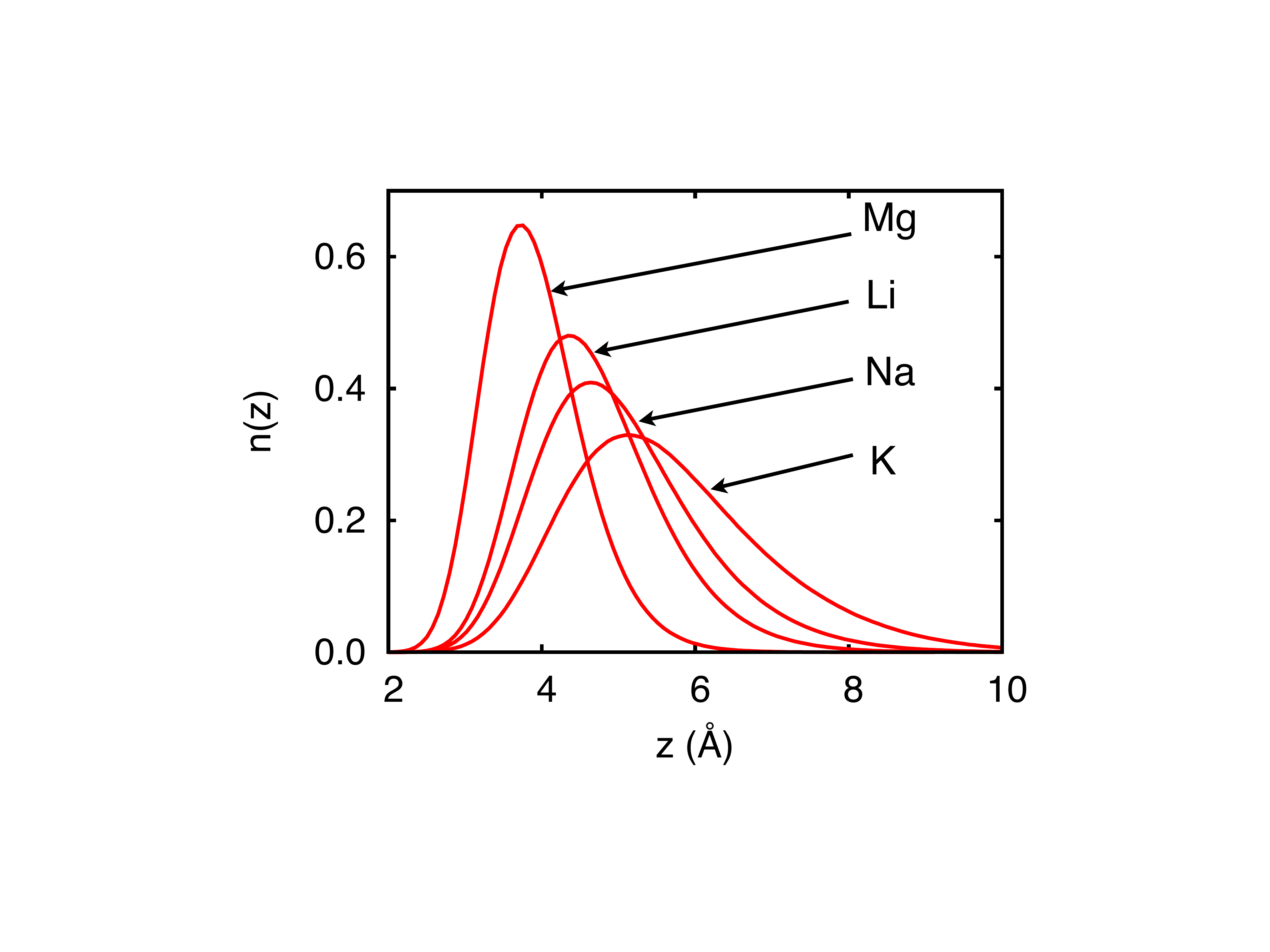}
\caption{(Color online) $^3$He density profile $n(z)$ (arbitrary units) in the direction perpendicular to the surface, on a Mg, Li, Na and K substrate. The height of the main peak is greater the stronger the substrate (see text). Results shown are for a coverage $\theta$=0.03 \AA$^{-2}$.}
\label{f2}
\end{figure}
\\ \indent
In order  to gain  insight into the physical properties of the adsorbed liquid films, we compute different structural properties, using extrapolated DMC estimators, typically adopted in DMC to obtain ground state expectation values of quantities other than the energy \cite{extra}. Although they are not numerically exact, as they retain some of the bias of the starting trial wave function, these estimators are usually fairly reliable, especially if, as observed in this work, the difference between the variational and ``mixed" estimate is small.
\\ \indent
Fig. \ref{f2} shows the integrated $^3$He density profile $n(z)\equiv \int dx dy\ \rho(x,y,z)$, where $\rho(x,y,z)$ is the 3D $^3$He density, for the substrates considered here. The height of the main peak is greater the stronger the substrate. Results shown are for a coverage $\theta=0.03$ \AA$^{-2}$, but we have found this quantity to be nearly independent of $\theta$, i.e., the density profile is almost entirely determined by the interaction of the helium atoms with the substrate, not surprisingly given the relatively large average planar interatomic separation in the range of coverage considered here. The difference in the spread of the $^3$He atomic wave function in the $z$ direction between the Mg substrate and the other ones, as well as the greater distance from the substrate, is  evident.  In particular, for the two substrates for which a quasi-2D liquid exists, the rms average excursion of each atom from its most probable distance from the substrate is of the order of a third of the average planar interparticle distance.
\begin{figure}[h]
\includegraphics[scale=0.36]{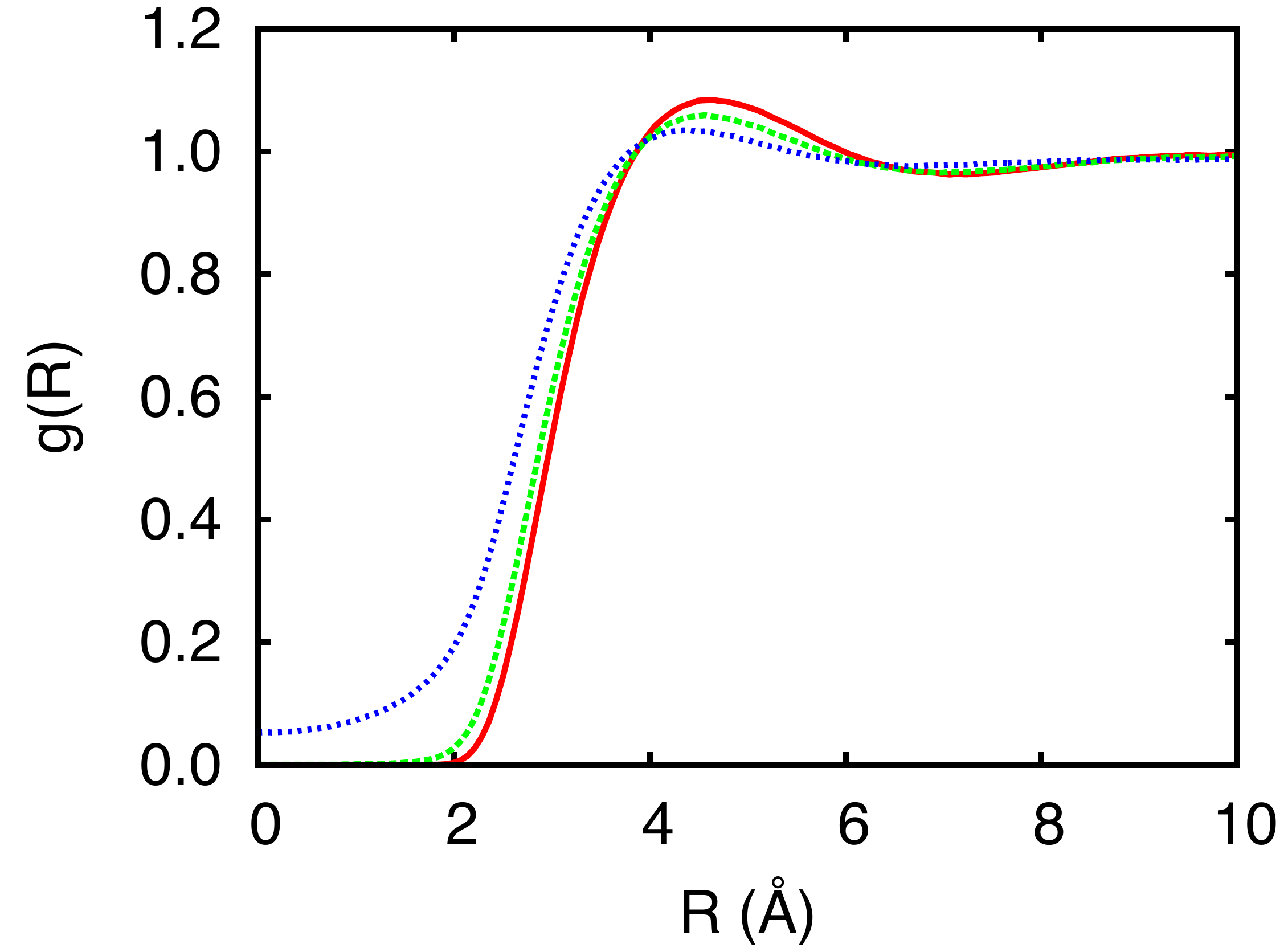}
\caption{(Color online)  Reduced pair correlation function $g(R)$ (Eq. (\ref {spat})), for $^3$He in 2D (solid line), on a Mg substrate (dashed line) and on a K substrate (dotted line), for a coverage $\theta=0.03$ \AA$^{-2}$. On a K substrate, $g(0)$ is finite. }
\label{f3}
\end{figure}
\\ \indent
With such a large spread, one might wonder to what extent the 2D characterization might be appropriate for such a system. A useful quantity is 
the angularly averaged, ``reduced" pair correlation function 
$g(R)$, with $R=\sqrt{x^2+y^2}$ and 
\begin{equation}
g(x,y) = {1\over  A \theta^2} \
\int dx^\prime dy^\prime \ n (x+x^\prime,y+y^\prime) \ n(x^\prime,y^\prime)
\label{spat}
\end{equation}
with $n(x,y)=\int dz \ \rho(x,y,z)$. The  more two-dimensional an adsorbed film, the more closely $g(R)$ mimics the pair correlation function of a strictly two-dimensional system of the same coverage. Fig. 3 shows results for $g(R)$ computed in the strictly 2D case, on a Mg substrate and on a K substrate, at a coverage $\theta=0.03$ \AA$^{-2}$. The results for the 2D case and on Mg are very similar, the main peak being slightly higher in 2D, and $^3$He atoms being able to come to slightly closer planar distances on a Mg substrate.  In other words the physics on a Mg substrate approaches fairly closely the 2D limit. On the other hand, the result on a K substrate is qualitatively different, the main feature being a finite value of $g(R)$ in the $R\to 0$ limit. This is a direct consequence of the large zero-point motion in the direction perpendicular to the substrate, whereby $^3$He atoms can occasionally be ``on top" on each other. For this reason, the qualified ``quasi" seems particularly appropriate when referring to this liquid phase as 2D.
\\ \indent
Thus, as also suggested in Ref. \cite{kr}, the main physical result of the large zero-point excursion is that of effectively softening the short-distance repulsive core of the interaction between two $^3$He atoms. By rendering the pair potential less repulsive, the substrate allows for the formation of a liquid.
This confirms an early prediction by Carraro and Cole \cite{carraro} that $^3$He can wet substrates that are not wetted by $^4$He, in this case Cs.
Conversely, and in some respects counter-intuitively, on a stronger, more attractive substrate such as Mg, or graphite, the physics of the system closely approaches the 2D limit, in which no liquid is expected to exist. It should be noted that the suggestion that zero-point motion in the transverse direction may stabilize a liquid phase of $^3$He was made by Brami {\it et al.} \cite{brami} for a (smooth model of a) graphite substrate, on which such an effect  actually does {\it not} occur, as mentioned above.
\\ \indent
We now discuss the main sources of systematic error of our calculation, in order to assess their expected effect on the physical conclusions outlined above. These are, for the energy, the finite size of the simulated system, the DMC time step error, the bias due to a finite walker population and the FNA, 
based on the  nodal structure given by Eq. (\ref{one}) \cite{nota_mix}. 
An important general remark applies to each of the systematic errors, namely that they cause an upward shift on the energy, of magnitude increasing with coverage (this behavior as a function of the coverage is an empirical result for the time step error, and an expected feature of all other sources of bias). 
We begin with finite-size effects. As mentioned above, the results shown here are for a system comprising $N$=26 particles; we have also carried out specific simulations with$N$=42 and 58 particles, obtaining results consistent with those shown here.
The leading finite-size correction to the energy is given by the 2D Fermi energy contribution, which is proportional to the areal density and greater for the $N=26$ than for the infinite system, as we ascertained by comparing results for different systems sizes. Morever, the population control bias \cite{noialtri} and the fixed-node error are both positive, and both vanish in the $\theta\to 0$ limit (where the trial wave function becomes exact). \\ \indent
Thus, all of these errors combined are not expected to affect the main conclusion of the study, 
which is the presence of a local minimum in the $e(\theta)$ curve on Na and K sustrates. If anything, it is possible that one such minimum may exist for Li as well, as suggested by the results shown with open symbols in Figure \ref{f1}. These  energies are obtained by applying  corrections to the simulation results, by estimating the magnitude of all systematic errors; in particular, the generally unknown magnitude of the  fixed-node 
error is assessed by looking at the difference between fixed-node and transient estimate \cite{nava} for the purely 2D case. The correction suggests that a liquid phase may exist on Li as well, with an equilibrium coverage of approximately 0.015 \AA$^{-2}$ (a 
similar correction would not change quantitatively the physical conclusions for all other substrates). 
\\ \indent
An interesting question is at what temperature the thermodynamically stable $^3$He film should turn superfluid. The methodology that we have adopted in this work does not allow us to explore this issue directly by simulation,  due to the sign problem. An order of magnitude estimate might be obtained by computing the binding energy of a pair of $^3$He atoms in the presence of the substrate, on the assumption that superfluidity of a Fermi system should  be underlain by the 
formation of bound pairs of atoms of opposite spins \cite{randeria,randeria2}. In Ref. \cite{kr}, the claim is made that the binding energy, which is 20 $\mu$K in 2D, 
could be as high as $\sim$ 10 mK in the vicinity of a weak substrate. Thus, an order of magnitude estimate of the superfluid transition temperature is of the order of 1 mK.
\\ \indent
Summarizing, we have carried out QMC simulations yielding robust numerical evidence to the effect that $^3$He will form a  stable, quasi-2D liquid phase at $T$=0 on Na, K, Rb and Cs substrates, and possibly on Li as well. The formation of the thermodynamically stable liquid phase is a consequence of the large atomic zero-point motion in the direction perpendicular to the substrate, which has the effect of softening the repulsive part of the helium interatomic potential at short distances. Based on this effect alone, no thin film is predicted to form on stronger substrates, such as Mg and graphite. The predictions made in this work appear to be experimentally testable, given the wealth of investigative work carried out over the past two decades, aimed at characterizing the physics of $^4$He films adsorbed on alkali substrates \cite {nacher91,rutledge92,ketola92,bigelow92,taborek94,mistura94,demolder95,hallock95,roller95,wyatt95,klier95,ross96,phillips98,moreau98,vancleve,taborek09}.
\\ \indent
	The authors wish to acknowledge useful discussions with Milton W. Cole and N. V. Prokof'ev.


\begin{thebibliography}{99}
\bibitem{note2}
The values in K are $-22.27$ on Mg, $-9.81$ on Li, $-6.41$ on Na, $-3.74$ on K, $-3.31$ on Rb and $-3.14$ on Cs.
\bibitem{nota_mix} 
For quantities other than the energy, the use of the extrapolated estimators is an additional source of error.
\end{thebibliography}
\end{document}